\documentclass[runningheads]{llncs}
\usepackage[T1]{fontenc}

\usepackage{graphicx}
\usepackage{booktabs}
\begin{document}
\title{From Complaint Narratives to Monetary Relief: A Hybrid Machine Learning Framework for CFPB Consumer Complaints}
\titlerunning{Hybrid ML for CFPB Monetary Relief Prediction}
\author{Zhuoer Wang\thanks{All authors contributed equally to this work.}\inst{1}\orcidID{0009-0005-3596-1391} \and
Sizhen Zhu\inst{2}\orcidID{0009-0000-5366-1257} \and
Xiongyu Chen\inst{3}\orcidID{0009-0006-6542-0147}}
\authorrunning{Z. Wang et al.}
\institute{University of Illinois Urbana-Champaign \email{zhuoerwang709@gmail.com}\and
North Carolina State University \email{szhu13@alumni.ncsu.edu} \and
Carnegie Mellon University \email{chen22mail@gmail.com}}
\maketitle

\begin{abstract}
Consumer financial complaints provide a valuable source of information for identifying service failures, dispute frictions, and operational deficiencies in consumer-facing financial institutions. This paper proposes a hybrid machine learning framework for predicting monetary relief outcomes using Consumer Financial Protection Bureau complaint data. We formulate the task as an imbalanced binary classification problem, where complaints closed with monetary relief are treated as compensable outcomes. The proposed framework integrates multiple sources of predictive information, including complaint narrative text, LDA-based topic representations, interpretable text-engineered features, and structured categorical attributes such as company and state. An XGBoost classifier is trained using a temporal train-test split, with earlier complaints used for model development and more recent complaints reserved for out-of-sample evaluation. Compared with a TF-IDF baseline, the proposed framework substantially improves predictive performance, increasing AUC-ROC from 0.69 to 0.78 and improving PR-AUC under class imbalance. Feature importance analysis shows that textual signals, latent complaint topics, and company identity all contribute meaningful predictive information. In particular, company-level effects reveal systematic variation in complaint resolution patterns across financial institutions. These findings suggest that consumer complaint narratives can serve as alternative data for monitoring consumer harm, identifying firm-level operational weaknesses, and supporting early-stage risk surveillance in consumer finance.


\end{abstract}

\newpage

\section{Introduction}

Consumer-facing financial services rely on complex operational processes for transaction handling, account management, and dispute resolution. When these processes break down through service errors, processing mistakes, or other frictions, the result is often a consumer complaint, and in a meaningful share of cases, monetary remediation. Identifying which complaints are likely to end in financial relief is valuable both to regulators seeking to monitor consumer harm and to firms seeking to understand and reduce their own operational failures.

Consumer complaint data provide a valuable yet underutilized source of information on such issues. In particular, the Consumer Financial Protection Bureau (CFPB) complaint database contains detailed narratives describing consumer experiences with financial products and services. These narratives capture aspects of service delivery and dispute processes that are not observable in traditional structured financial data.

In this paper, we build a model that predicts whether a complaint will be resolved with monetary relief. We treat complaints closed with monetary relief as a proxy for compensable outcomes, that is, cases where a service error or operational friction was significant enough to warrant financial remediation. Because only about 15\% of complaints fall into this category, the prediction task is characterized by substantial class imbalance, which we account for throughout training and evaluation.

We develop a hybrid framework combining topic modeling, text-based feature engineering, and gradient boosting to estimate the likelihood of  monetary relief outcomes, allowing the model to draw on both latent semantic structure in the narratives and structured categorical attributes such as company identity and state. Our results show that complaint narratives contain significant predictive signals, achieving an AUC-ROC of 0.78 under class imbalance, and the model generalizes to a broader set of complaint categories when credit card complaints are added. Explainability analyses further reveal that both semantic features and latent topics contribute to model performance.

Importantly, we find that company identity is a highly significant predictor, indicating systematic differences in complaint patterns and outcomes across financial institutions. This suggests that complaint data may provide observable signals of firm-level heterogeneity in service processes and potential operational deficiencies. This dual relevance makes the framework useful from two perspectives: regulators can use it as an early-stage alerting mechanism to flag complaints and firms associated with potential monetary impact, while companies can use it to detect operational deficiencies in their own service processes. 

Overall, our findings highlight the potential of complaint narratives as alternative data for monitoring consumer-facing financial services and identifying patterns associated with compensable outcomes.

\section{Related Work}

Prior research on the CFPB Consumer Complaint Database falls into several strands, 
including topic discovery, complaint classification, prediction of company responses and outcomes, and the role of large language models. We review each in turn and situate our work relative to them.

\textbf{Topic modeling and theme discovery. }A first line of work uses unsupervised methods to surface latent structure in complaint narratives. Bastani et al. apply Latent Dirichlet Allocation to the CFPB database to identify recurring themes and track how complaint topics evolve over product categories and time \cite{LDA-CFPB}. Such methods provide interpretable summaries of large narrative corpora but stop short of predicting outcomes. More recent work has examined the reliability and stability of topic modeling itself, raising methodological cautions about how reproducibly topics can be recovered \cite{topic-reliability}. We build directly on this strand by incorporating LDA-derived topic representations as features rather than as a standalone descriptive tool.

\textbf{Complaint classification.} A second strand treats complaints as a supervised classification problem, typically predicting product or issue categories. Approaches range from classical machine learning pipelines \cite{ml-complaints} to specialized deep architectures such as a two-stage residual one-dimensional convolutional network designed for the CFPB narratives \cite{tsr1dcnn}. A growing body of work applies large language models to this task: comparative studies report that certain LLMs achieve strong zero-shot classification performance relative to others \cite{claude-zeroshot}, and retrieval-augmented, relevancy-driven few-shot prompting has been shown to improve LLM classification beyond the zero-shot baseline \cite{rag-fewshot}. These studies focus on categorizing complaints rather than predicting their financial resolution.

\textbf{Predicting company responses and resolution outcomes.} Closest to our objective is work that predicts how complaints are resolved. The Consumer Feedback Insight and Prediction Platform uses machine learning on the full CFPB database to predict both the timeliness and the nature of company responses, including whether a complaint is closed with relief, and pairs this with LDA-based theme discovery for consumers and regulators \cite{predictive-cfpb}. Related work develops human-experience-trained classifiers and expert-refined synthetic data generation to better capture the nuanced linguistic cues relevant to relief eligibility \cite{eval-metrics}. A separate body of work studies anomaly detection and fraud risk in financial narratives and transactions \cite{llm-fraud}, and economic studies use CFPB data to quantify consumer exposure to fraud and scams \cite{fraud-exposure}. Our work shares the goal of predicting compensable outcomes but differs in combining LDA topics, LLM-derived semantic features, and structured categorical attributes within a single gradient-boosted model, and in adopting a temporal train/test split to reflect a realistic deployment setting.

\textbf{Large language models and consumer communication.} A final strand examines LLMs as actors in the complaint process rather than as analytical tools. Sang and Bae provide observational and experimental evidence that consumers increasingly use ChatGPT to draft complaints, that AI-assisted complaints differ measurably in linguistic features, and that these complaints are more likely to elicit relief from firms \cite{llm-adoption}. Broader work surveys generative AI for the classification, summarization, and response generation of consumer communications \cite{genai-comms}. From a regulatory perspective, recent studies show that public disclosure of CFPB complaints disciplines banks through depositor responses and use LLMs to extract qualitative topic dimensions from narratives \cite{disciplining-banks}, and others propose governance frameworks for AI-driven fraud detection that align model development and monitoring with regulatory requirements \cite{governance-framework}. These works motivate the dual regulatory and firm-level relevance of complaint analysis, which our framework operationalizes by treating company identity as an explicit, interpretable feature.

In summary, prior work has separately explored topic discovery, classification, outcome prediction, and the influence of LLMs on CFPB complaints. Our contribution unifies semantic and structured signals in a single predictive framework for compensable outcomes, emphasizes interpretability through feature importance, and surfaces systematic cross-firm heterogeneity relevant to both regulators and financial institutions.

\section{Data}
\subsection{Data Characteristics}

The data analysis in this article uses Consumer Financial Protection Bureau (CFPB) complaint data containing 176,942 complaint records and 18 variables related to consumer banking complaints. The dataset focuses on complaints associated with checking and savings account products and includes information on complaint characteristics, reporting channels, geographic location, involved financial institutions, and company responses.

According to the modeling target, the following three key variables are selected for further complaint distribution learning, including: 

\begin{itemize}
    \item \texttt{Timely Response}: 99.0\% of complaints received a timely response (175,206 cases), while only 1.0\% were classified as untimely (1,736 cases).
    \item \texttt{Company Response to Consumer}: Most complaints were closed with explanation (140,703 cases, 79.5\%), followed by monetary relief (26,445 cases, 14.9\%) and non-monetary relief (9,661 cases, 5.5\%). Untimely responses and in-progress cases represent a very small proportion of observations.
    \item \texttt{Company Public Response}: Most companies either chose not to provide a public response after responding to the consumer and CFPB or provided standardized response categories, resulting in a highly skewed distribution across response classes. 
\end{itemize}

Taking a further check of the data, the following variables are being evaluated to gain us a better understanding on how those variables could potentially affect the complaint responses above:

\begin{itemize}
    \item \texttt{Complaint Issue Categories}: Complaint issues are concentrated in a small number of categories. The most common issue is Managing an Account (96,156 complaints), accounting for more than half of all records. Other frequent issues include Closing an Account, Problems with Account Charges, and Low Funds-Related Problems. At the sub-issue level, categories such as Deposits and Withdrawals, Debit/ATM Card Problems, and Unauthorized Transactions are particularly prevalent.
    \item \texttt{Company Concentration}: The dataset is highly concentrated among a limited number of financial institutions. Large national banks account for a substantial share of complaints, including Wells Fargo, JPMorgan Chase, Bank of America, Navy Federal Credit Union, and Capital One. To facilitate analysis, the largest institutions were analyzed individually while smaller organizations were grouped into an "Others" category.
\end{itemize}

\begin{figure}
\includegraphics[width=1\textwidth]{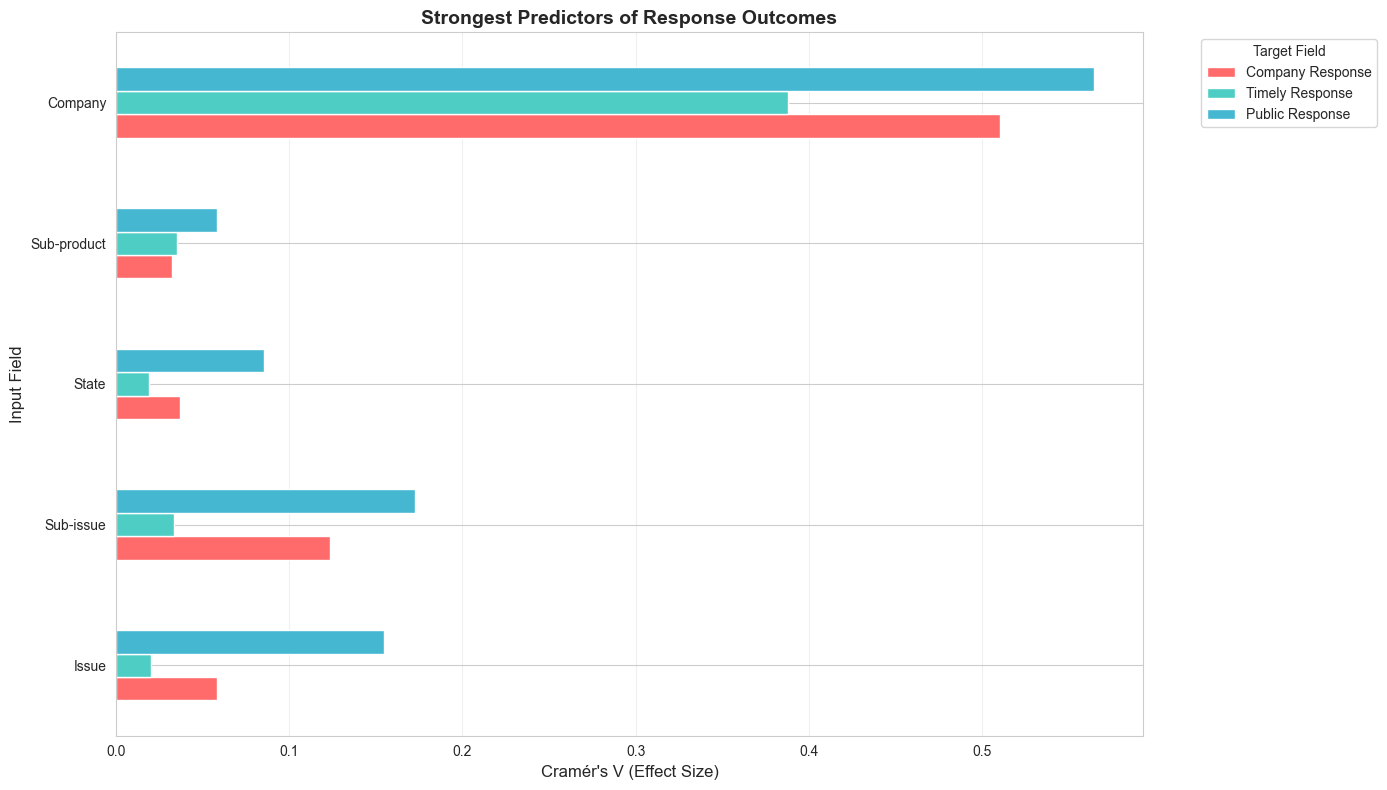}
\caption{COMPARATIVE ANALYSIS: Strongest Predictors Across All Target Fields}
\label{fig2}
\end{figure}

The following section exploits the potential relationship between predictor and the complaint responses variables. Chi-square tests and Cramér’s V statistics were used to evaluate associations between complaint attributes and three target variables: Company Response to Consumer, Timely Response, and Company Public Response. The results indicate that Company is consistently the strongest predictor across all three outcomes Table-\ref{tab:cramersv} and Figure-\ref{fig2}:

\begin{table}[htbp]
\centering
\caption{Strongest categorical predictor for each target variable by Cramér's V.}\label{tab:cramersv}
\begin{tabular}{llc}
\hline
Target Variable & Strongest Predictor & Cramér's V \\
\hline
Company Response to Consumer & Company & 0.510 \\
Timely Response & Company & 0.388 \\
Company Public Response & Company & 0.565 \\
\hline
\end{tabular}
\end{table}

According to conventional interpretation guidelines, these values indicate moderate to strong associations, suggesting that response behavior varies substantially across financial institutions. Complaint content variables also exhibit predictive value. Sub-issue demonstrates weak-to-moderate associations with both Company Response to Consumer (Cramér's V = 0.124) and Company Public Response (Cramér's V = 0.173), while broader Issue categories show weaker but statistically significant relationships. Geographic location (State) and account type (Sub-product) have comparatively limited explanatory power.

Among the three target variables, whether the company eventually “reply with monetary relief” is our most interested in topic. As shown in Figure-\ref{fig3} below, we observed that the ratio of “closed with monetary relief” has gradually decreased over the year. 

\begin{figure}
\centering
\includegraphics[width=1\textwidth]{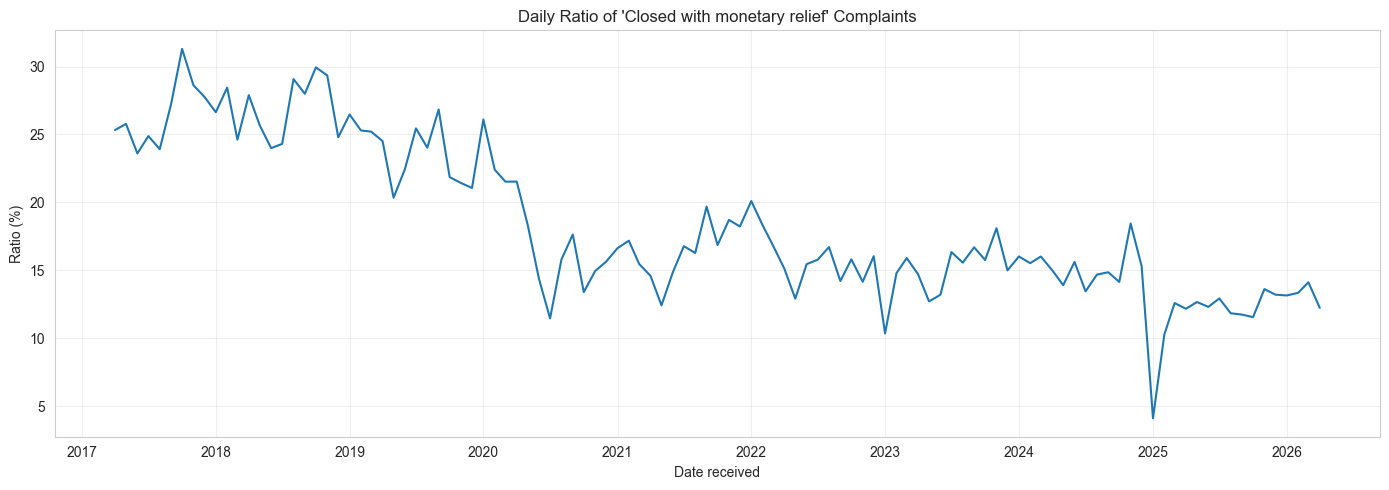}
\caption{Daily Ratio of 'Closed with monetary relief' Complaints}
\label{fig3}
\end{figure}

A further exploration of the relationship between key individual variables and \% ratio of “reply with monetary relief” by time is conducted below, to better understanding which predictor could be the key driver of the dropped trend. 

\subsubsection{Company vs. \% of reply with monetary relief}
Figure-\ref{fig4} below illustrates the daily proportion of complaints closed with monetary relief across major financial institutions. Several notable patterns emerge. First, substantial variation exists both across institutions and over time, indicating that monetary relief practices are not uniform throughout the study period.

\begin{figure}
\centering
\includegraphics[width=1\textwidth]{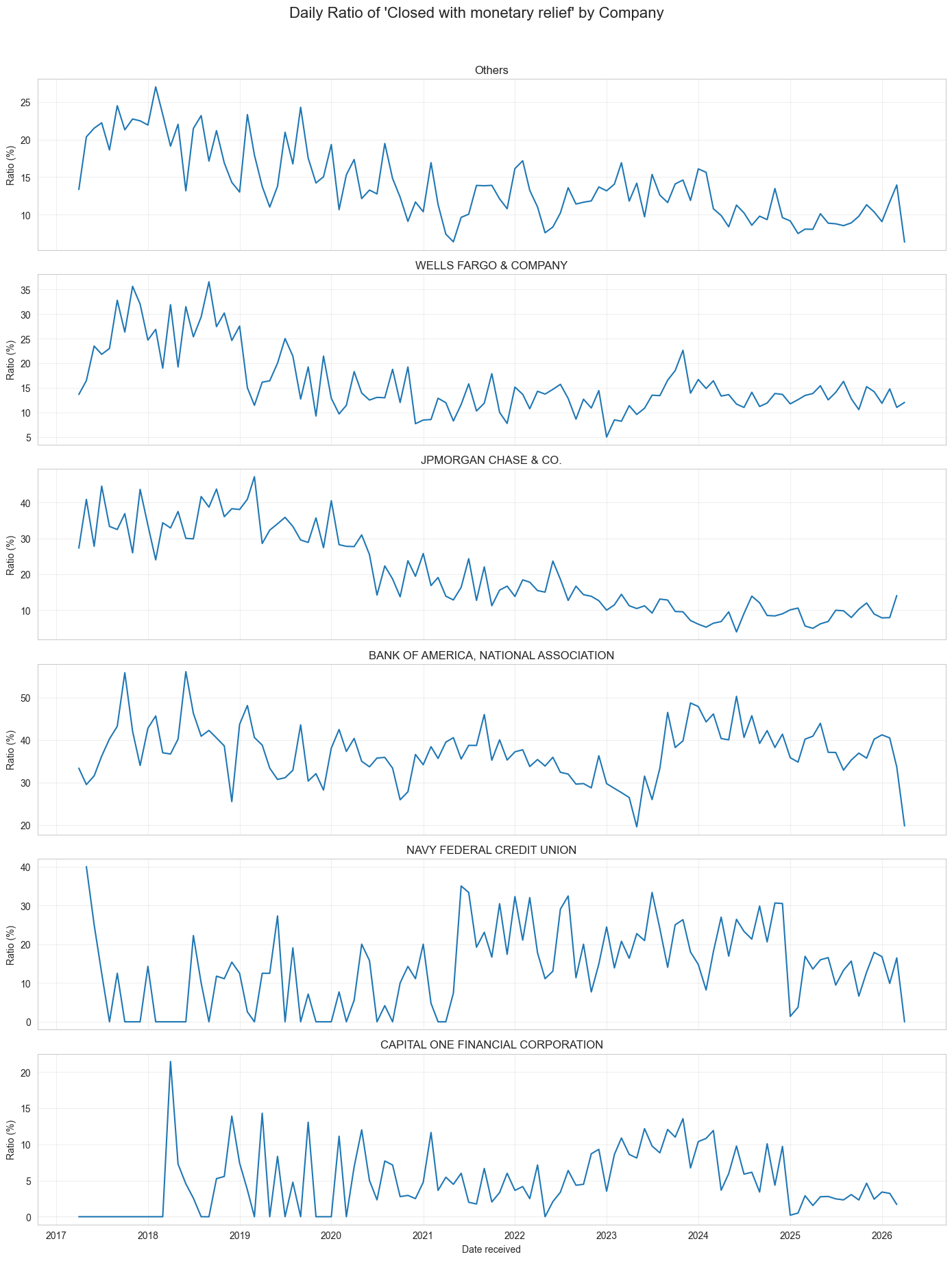}
\caption{Daily Ratio of 'Closed with monetary relief' Complaints by Company}
\label{fig4}
\end{figure}

Most institutions exhibit a general decline in the proportion of complaints resulting in monetary relief from 2018 to 2026. This trend is particularly pronounced for JPMorgan Chase, where the ratio decreases from approximately 35–45\% in the early years to below 15\% in recent periods. Wells Fargo also shows a downward trend, although the decline is less steep and is accompanied by periods of recovery.

Bank of America maintains comparatively high monetary relief rates throughout the observation period, typically ranging between 30\% and 50\%, suggesting a different complaint resolution pattern compared with other institutions. In contrast, Capital One consistently records the lowest monetary relief rates, generally below 15\% and declining to below 5\% in later years.

The Navy Federal Credit Union series displays considerably higher volatility than other institutions, with frequent fluctuations and several periods of near-zero monetary relief rates. This instability may reflect lower complaint volumes, greater sensitivity to individual cases, or changes in complaint handling policies over time.

Overall, the observed temporal trends suggest that both institutional characteristics and time-dependent factors influence complaint resolution outcomes. The variation across companies supports the earlier finding that company identity is one of the strongest predictors of consumer response outcomes. Additionally, the changing patterns over time indicate potential non-stationarity in the data, reinforcing the importance of incorporating temporal information into predictive modeling and model validation procedures.

\subsubsection{Issue vs. \% of reply with monetary relief}
\begin{figure}
\centering
\includegraphics[width=1\textwidth]{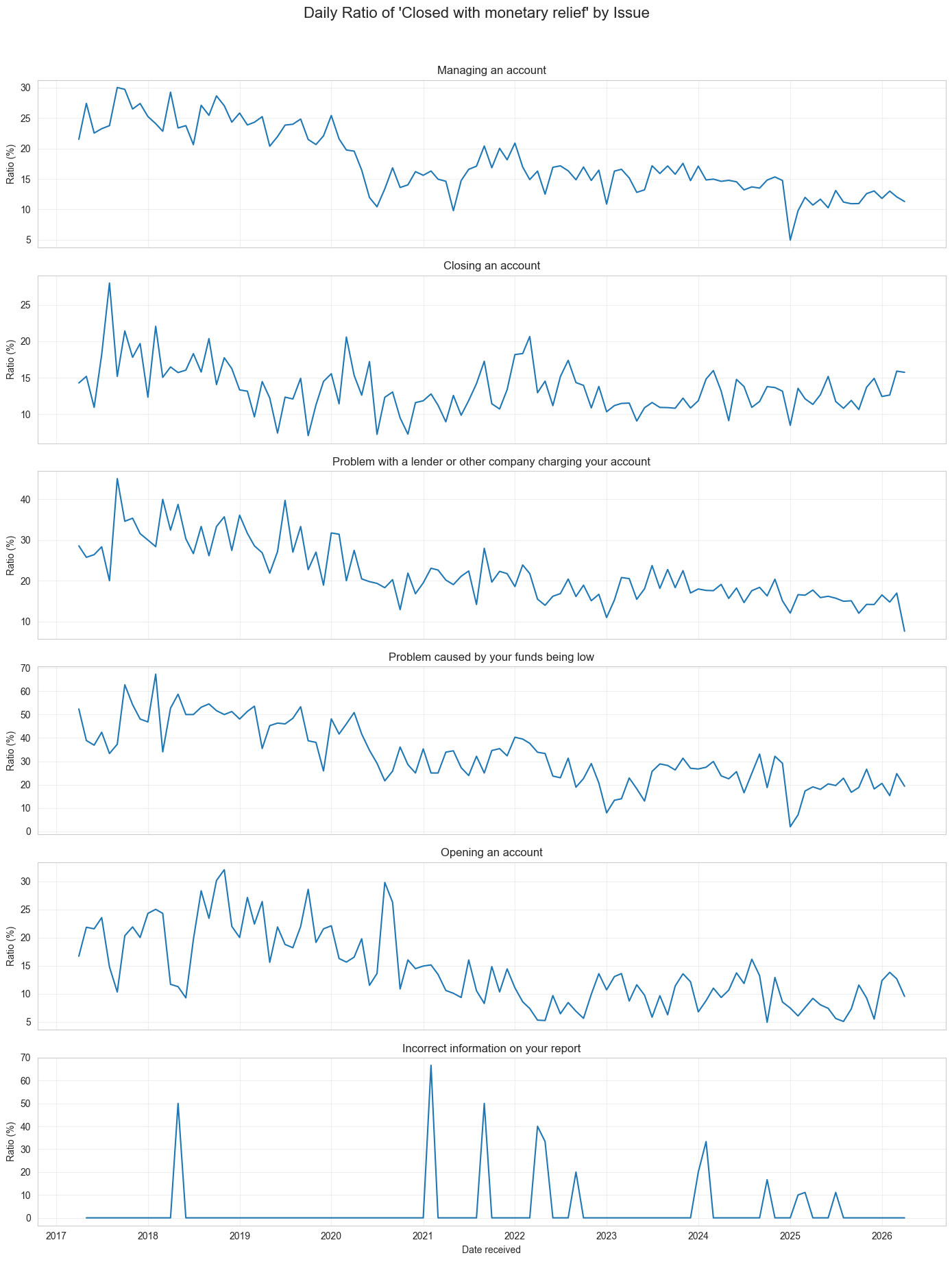}
\caption{Daily Ratio of 'Closed with monetary relief' Complaints by Issues}
\label{fig5}
\end{figure}

Figure-\ref{fig5} presents the daily proportion of complaints closed with monetary relief across the most common complaint issue categories. Similar to the company-level analysis, substantial variation is observed both across complaint types and over time.
Among the major complaint categories, Problems Caused by Your Funds Being Low consistently exhibits the highest monetary relief rates, frequently ranging between 20\% and 50\% during the early years of the dataset. Although the proportion declines over time, it remains higher than most other issue categories throughout the observation period. This suggests that complaints involving overdraft fees, insufficient funds, and related account charges are more likely to result in financial compensation.

Managing an Account, the most prevalent complaint category, shows a gradual decline in monetary relief rates from approximately 25–30\% in earlier years to around 10–15\% in more recent periods. Given the large volume of complaints in this category, this downward trend contributes significantly to the overall reduction in monetary relief observed in the dataset.

Complaints related to Problems with a Lender or Other Company Charging Your Account and Closing an Account display moderate monetary relief rates and relatively stable patterns over time. While both categories experience fluctuations, their long-term trends suggest a gradual reduction in the likelihood of monetary compensation.
The Opening an Account category exhibits considerable volatility, particularly during the earlier years of the study period. Monetary relief rates decline substantially after 2020, stabilizing at relatively low levels thereafter. This pattern may reflect changes in complaint composition, institutional policies, or regulatory practices affecting account-opening disputes.

A distinct pattern is observed for Incorrect Information on Your Report. Unlike other complaint categories, this issue type is characterized by infrequent but extreme spikes in monetary relief rates, including several periods exceeding 50\%. However, these spikes are separated by long intervals with few or no observations, suggesting that the category contains a relatively small number of complaints and is therefore more sensitive to individual cases.

Overall, the figure indicates that complaint issue type influences both the level and temporal evolution of monetary relief outcomes. While company identity remains the strongest predictor of complaint resolution outcomes, the observed differences across issue categories provide additional evidence that complaint content contributes meaningful explanatory information. This observation is consistent with the Cramér's V analysis, which identified statistically significant associations between Issue/Sub-issue variables and Company Response to Consumer outcomes.

\subsubsection{Sub-Issue vs. \% of reply with monetary relief}
\begin{figure}
\centering
\includegraphics[width=1\textwidth]{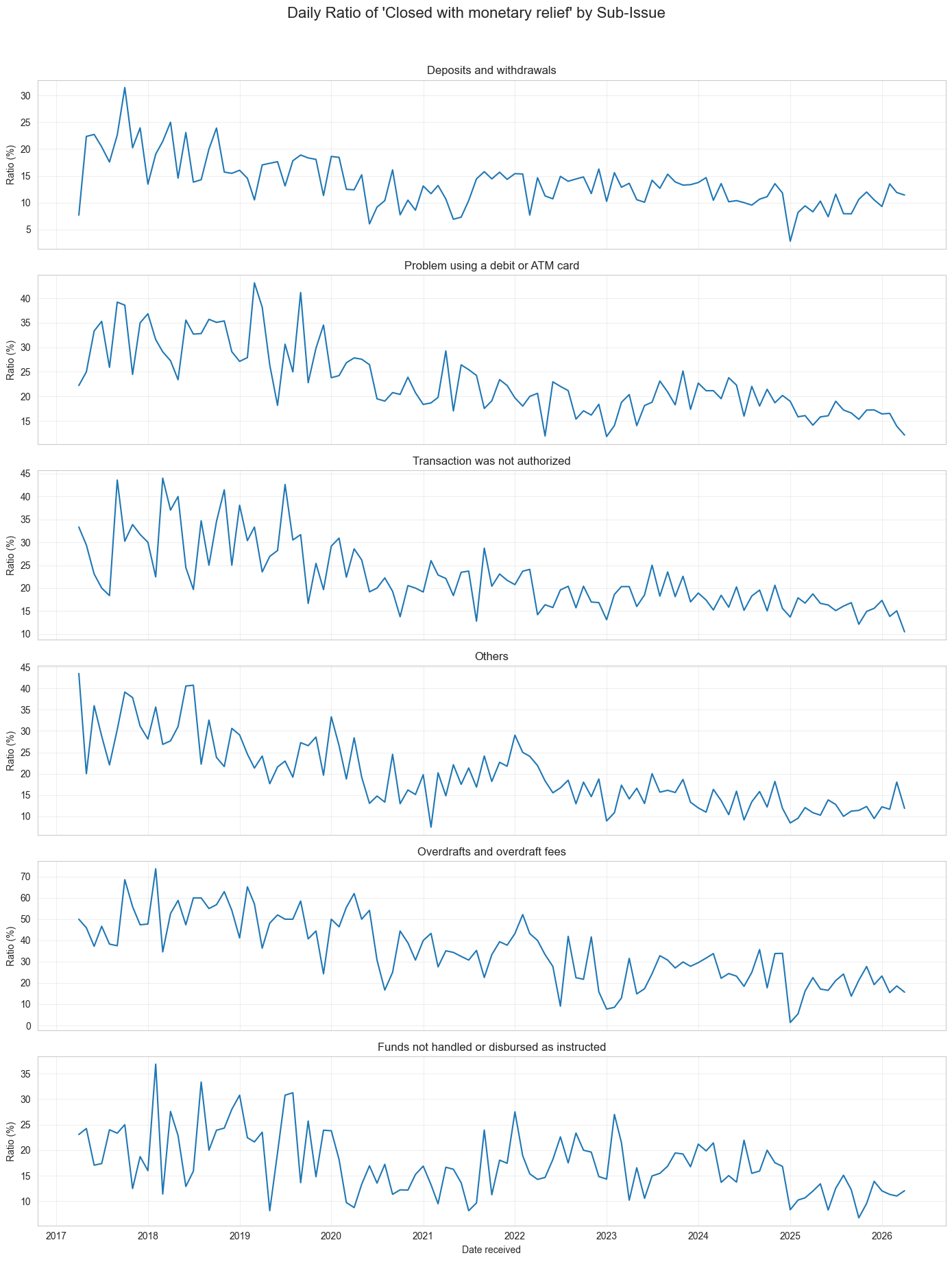}
\caption{Daily Ratio of 'Closed with monetary relief' Complaints by Issues}
\label{fig6}
\end{figure}

Figure-\ref{fig6} illustrates the daily proportion of complaints closed with monetary relief for the most frequently occurring sub-issue categories. Compared with the broader issue-level analysis, the sub-issue trends reveal more granular differences in complaint resolution outcomes and provide additional evidence that specific complaint circumstances influence the likelihood of monetary compensation.

Among the major sub-issues, Overdrafts and Overdraft Fees consistently exhibits the highest monetary relief rates throughout the study period. Although the proportion declines over time, it remains substantially higher than most other categories, often ranging between 20\% and 50\%. This pattern suggests that complaints involving overdraft-related charges are more likely to result in financial remediation.

Deposits and Withdrawals, one of the most common sub-issues, shows a relatively stable but gradually declining trend. Monetary relief rates decrease from approximately 20–30\% during the earlier years to around 10–15\% in recent periods. Because this category represents a large share of complaints, it contributes significantly to the overall downward trend observed in monetary relief outcomes.

Complaints involving Problem Using a Debit or ATM Card and Transactions Not Authorized follow similar trajectories, with monetary relief rates steadily declining over time. Both categories maintain moderate compensation levels, generally ranging between 15\% and 35\%, suggesting that while monetary relief remains common for transaction-related disputes, its frequency has decreased in more recent years.

The Funds Not Handled or Disbursed as Instructed category exhibits greater volatility than the other major sub-issues. While monetary relief rates fluctuate considerably from period to period, the long-term trend also indicates a gradual decline. Such variability may reflect lower complaint volumes or greater heterogeneity in the underlying complaint circumstances.

The aggregated Others category demonstrates a broad downward trend with occasional short-term increases. Because this category combines numerous less frequent sub-issues, the observed pattern likely reflects a mixture of complaint types with differing resolution characteristics.

Overall, the figure suggests that monetary relief outcomes differ not only across broad complaint issues but also across specific sub-issues. The distinct temporal trajectories observed for individual sub-issues support the statistical findings from the association analysis, where Sub-issue exhibited stronger relationships with Company Response to Consumer outcomes than the higher-level Issue variable. These results indicate that detailed complaint classifications contain valuable information for predicting complaint resolution outcomes and should be retained as important features in subsequent modeling efforts.

Monthly trend analysis shows that response outcomes vary over time and across institutions. The proportion of complaints resulting in monetary relief changes across reporting periods and differs among major financial institutions. Similar temporal variation is observed for timely response rates and issue-specific complaint outcomes, indicating that complaint handling practices are not entirely stable over time.

\subsubsection{Implications for Modeling}
Several characteristics of the dataset are relevant for predictive modeling:

\begin{itemize}
    \item Severe class imbalance exists for some target variables, particularly Timely Response.
    \item Company-level effects dominate outcome variation, making institutional identity a critical predictor.
    \item Issue and sub-issue categories provide additional explanatory information, particularly for public and consumer-facing responses.
    \item Temporal variation suggests potential non-stationarity, indicating that time-based validation strategies may be appropriate.
    \item Most statistically significant relationships are driven by categorical variables, supporting the use of encoding methods and models capable of handling high-cardinality categorical features.
\end{itemize}

\subsection{Label Definition}
We use monetary relief as an observable indicator of cases where service deficiencies, processing errors, or dispute-related issues resulted in financial remediation. Although monetary relief does not constitute a direct measure of underlying service quality or consumer harm, it provides a consistent and economically meaningful signal of outcomes involving monetary compensation. Our objective is to identify complaints associated with a higher likelihood of compensable outcomes and to flag cases that may result in monetary impact, thereby enabling an early-stage alerting mechanism.

We defined a binary classification label based on the companies' response to each complaint. A complaint is labelled as 1 if the company's response is closed with monetary relief, and 0 otherwise. We observe that the data set is unbalanced, with only approximately 15\% of complaints labelled closed with monetary relief. While this imbalance reflects real-world scenarios, it requires careful consideration in training, evaluating, and interpreting.

\section{Methodology}
\subsection{Feature Engineering}

\subsubsection{LDA Data Mining on narratives – Two Layer Stratified NLP Model}

To further explore the data, especially to take advantage of more information from the customer's narratives, a two-layer stratified topic modeling framework was developed to capture multiple dimensions of customer complaint behavior and enrich the modeling features that are being covered in basic data. Rather than training a single global topic model across all complaint narratives, the approach trains separate topic models based on two important categorical dimensions:

\begin{itemize}
    \item \texttt{Issue Stratification}: models trained within each CFPB Issue category.
    \item \texttt{Company Stratification}: models trained within each financial institution.
\end{itemize}

The reason of choosing above two dimensions is that those are the most significant drivers which impact the modeling target "monetary relief". To be more specific, the model exploits the information with a two-layer structure. In layer 1, the NLP model was trained to reveal hidden topics from the two above dimensions. In layer 2, a Dual-Stratified Topic is determined and analyzed based on the cross-combination of the two layers. This design enables the extraction of both issue-specific and company-specific themes from complaint narratives and supports cross-dimension analysis of customer concerns.

\paragraph{Dimension 1 - Issue Stratification}

In the first stratification approach, complaint records were grouped according to their assigned Issue category. A separate topic model was trained for each Issue group using only complaint narratives associated with that Issue. For each issue category, ten issue-specific topics are generated,leading to a more granular-level information extraction. Especially, those topics capture more meaningful issue-specific patterns and customer concerns, which are customized for different types of issues.The Issue-stratified models reveal that complaint narratives are organized around a combination of:

\begin{itemize}
    \item Core banking operations (payments, deposits, account access)
    \item Fraud and dispute management
    \item Fees and overdraft concerns
    \item Customer support and communication issues
    \item Institution-specific complaint patterns embedded within broader Issue categories
\end{itemize}

\paragraph{Dimension 2 - Company Stratification}

A second set of topic models was trained using Company as the stratification dimension. Complaint narratives were grouped by financial institution, and a dedicated topic model was trained for each company. In this process, one topic model is trained per company. Within each company, ten topics are generated to capture company-related complaint patterns, which provide a better visibility of operational and company-service complaint themes. These models reveal themes that may be unique to a company's products, policies, customer service practices, communication style, or complaint handling procedures. Each company model is tuned to that institution’s complaint language and customer experience patterns. The topics therefore capture both generic banking themes that occur across institutions and company-specific operational or service characteristics. For example:

\begin{itemize}
    \item JPMorgan Chase \& Co.: Exhibits distinct topics related to account opening and closing, fraud disputes, ATM/deposit operations, and promotional account offers.
   
    \item Wells Fargo \& Company: Demonstrates strong topic clusters around complaint handling, fees and overdrafts, account management, and fraud-related disputes.
\end{itemize}

The Company-layer models reveal both universal banking concerns and institution-specific complaint behavior. Because each model is trained exclusively on complaints associated with a particular company, the resulting topics provide insight into:

\begin{itemize}
    \item Company-specific customer pain points
    \item Operational processes generating complaints
    \item Institution-level complaint trends
    \item Differences in how similar issues manifest across organizations
\end{itemize}

\paragraph{Summary - Dual-Stratified Approach}

After model training, both the Issue-based and Company-based models are applied to every complaint record. As a result, every complaint receives two independent topic interpretations:

\begin{itemize}
    \item An Issue-oriented topic assignment
    \item A Company-oriented topic assignment
\end{itemize}

These outputs are stored together in a unified analytical dataset, and being leveraged as the input parameters to enrich the drivers of the model. That is, this framework eventually enhances the hidden data feature reveal, improves analytical granularity, and provides a comprehensive view of CFPB complaint behavior across both issue and company dimensions.

\subsubsection{Text-based Feature Engineering}

We construct 14 interpretable text-based features from each consumer complaint narrative. These features are designed to capture complementary dimensions of the narrative, including its length and structure, linguistic style, financial specificity, documentation detail, and emotional emphasis. The features are extracted using deterministic rules, regular expressions, and simple statistical transformations, making them transparent and easy to interpret. Unlike high-dimensional sparse text representations, these handcrafted narrative features provide human-readable signals that can be directly examined through feature importance analysis in the predictive model.

\begin{table}
\centering
\caption{Definitions of text-based narrative features.}
\label{tab:feature_definitions}
\begin{tabular}{p{0.28\textwidth} p{0.62\textwidth}}
\hline
\textbf{Feature} & \textbf{Definition} \\
\hline
\texttt{word\_count} & Total number of whitespace-delimited tokens. \\
\texttt{char\_len} & Total character count of the narrative. \\
\texttt{sentence\_count} & Number of sentence-terminal punctuation sequences. \\
\texttt{paragraph\_count} & Number of double-newline-separated paragraphs. \\
\texttt{avg\_sentence\_len} & Ratio of \texttt{word\_count} to \texttt{sentence\_count}. \\
\texttt{lexical\_diversity} & Ratio of unique lowercase word types to total word tokens. \\
\texttt{avg\_word\_len} & Mean character length across all tokens. \\
\texttt{flesch\_kincaid} & Flesch Reading Ease score. \\
\texttt{dollar\_count} & Number of \texttt{\$}-prefixed numeric expressions. \\
\texttt{total\_dollar\_amount} & Sum of all parsed dollar values extracted from the narrative. \\
\texttt{max\_dollar\_amount} & Largest dollar value mentioned in the narrative. \\
\texttt{date\_count} & Number of date-formatted strings appearing in the narrative. \\
\texttt{caps\_ratio} & Proportion of fully capitalised tokens of length $> 1$, used as a proxy for written emphasis. \\
\texttt{exclamation\_count} & Number of \texttt{!} characters. \\
\hline
\end{tabular}
\end{table}

\subsubsection{Structured Categorical Features}

Company and State are incorporated as structured categorical features to capture geographic patterns as well as characteristics associated with different financial institutions. 

\subsection{Monetary Relief Prediction Model}

Given the characteristics of the dataset, monetary relief prediction is an imbalanced binary classification problem. We selected XGBoost as the primary predictive model due to its strong performance in imbalanced datasets and its ability to capture nonlinear relationships and interaction of features across heterogeneous predictors.

Our baseline is trained on TF-IDF representations of complaint narratives. We then incorporate an expanded feature set, including latent topic representations derived from LDA and semantically meaningful features selected through the text-based feature engineering process. The improved predictive performance observed in the enhanced models suggests that monetary relief outcomes are associated not only with presence or frequency of words and phrases but also with broader complaint themes and contextual characteristics embedded within consumer narratives.

To better simulate real-world applications with the latest real-time consumer complaints and predict newly submitted real-time complaints, instead of randomly splitting the training and testing set, we split the training and testing data based on the complaint submission date. The complaints are sorted by the date received with the earliest 80\% used for training and the most recent 20\% reserved for out-of-sample evaluation. 

\section{Result}
\subsection{Classification Performance}

Compared with the baseline TF-IDF approach relying solely on complaint narratives, the enhanced model with selected features achieved substantial improvements across multiple metrics. In particular, the AUC-ROC increased from 0.69 to 0.78, while the PR-AUC improved from 0.25 to 0.35. Given the imbalanced nature of monetary relief outcomes, the gain in precision-oriented metrics indicates a meaningful improvement in identifying complaints that are more likely to result in compensable resolutions.

\begin{table}[htbp]
\centering
\caption{Model Performance Comparison}
\label{tab:main_results}
\begin{tabular}{lccc}
\toprule
Model & AUC-ROC & PR-AUC & F1 Score \\
\midrule
TF-IDF (Baseline) & 0.6917 & 0.2523 & 0.3117 \\
All selected features & 0.7751 & 0.3522 & 0.3747\\
All selected features(extended dataset) & 0.7820 & 0.3937 & 0.3928\\
\bottomrule
\end{tabular}
\end{table}

\begin{figure}
\centering
\includegraphics[width=0.6\textwidth]{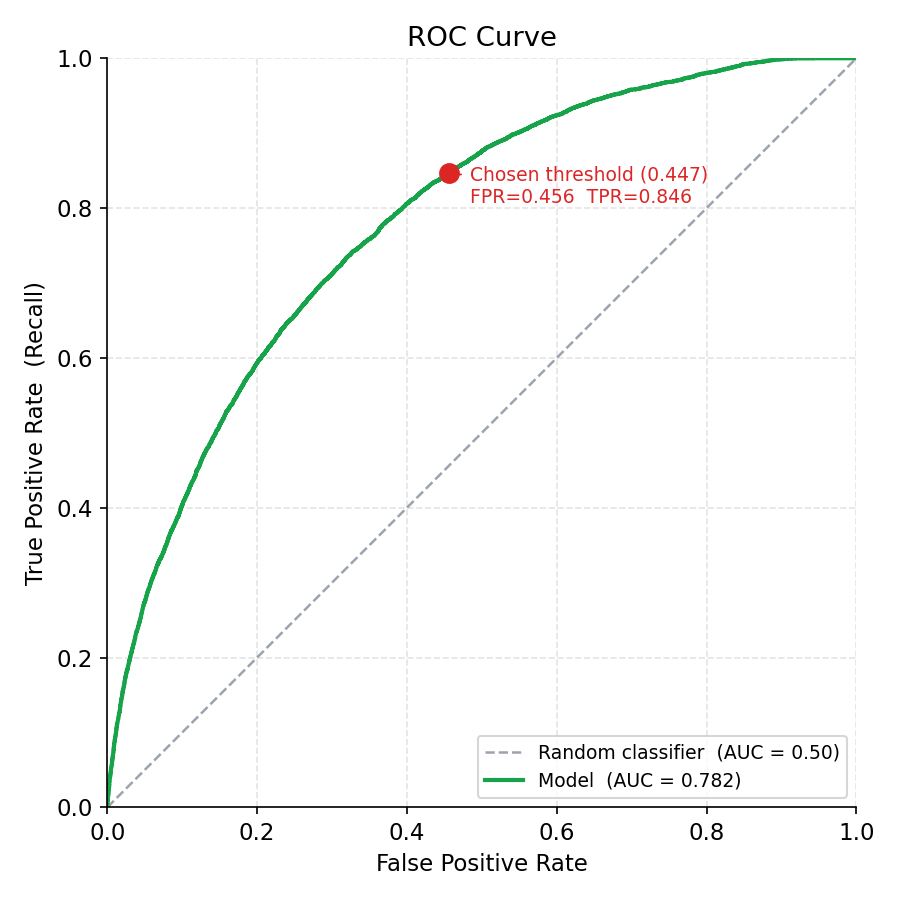}
\caption{ROC curve for the XGBoost model trained on the full feature set, with an emphasis on achieving high recall.}
\label{fig1}
\end{figure}

The observed performance gain suggests that monetary relief outcomes are influenced not only by the occurrence and frequency of individual words and phrases captured by TF-IDF representations, but also by higher-level semantic structures and recurring contextual themes embedded within complaint narratives. These findings indicate that latent topics, the constructed semantic features, and the financial entity provide complementary predictive information beyond traditional bag-of-words approaches.

Figure~\ref{fig1} shows the ROC curve of the enhanced XGBoost model. The model demonstrates strong discriminative ability with a target on high recall, reflecting the purpose of minimizing missed cases among complaints that ultimately result in monetary relief. Such identification may support regulators in monitoring consumer harm and help firms recognise and address underlying operational failures.

Additionally, we evaluate the proposed model on an extended dataset that also includes credit card-related complaints. The model achieves an AUC-ROC of 0.7820, suggesting that the selected features generalise effectively across a broader category of financial consumer complaints and capture patterns that are not restricted to a single complaint domain.

\subsection{Selected Significant Features and Interpretation}
The final model retained the top 200 features with non-zero importance scores. Within the 100 highest-ranked predictors, 33 were derived from text-based representations, 13 corresponded to latent complaint topics identified through LDA, and 54 represented structured categorical attributes, including company and state information.

The distribution of important features proved the value of integrating heterogeneous sources of information for monetary relief prediction. While text-based features from complaint narratives remained influential, topic-based and structured features also provide valuable information. 

Our analysis shows that company identity is one of the most significant features. This result indicates that complaint outcomes are not solely determined by the descriptions and topics of individual cases, but also exhibit systematic variation across financial institutions.

This variation across financial institutions suggests that consumer complaint data encode meaningful signals related to differences in service quality, operational processes, and dispute handling practices. As such, consumer complaints as a type of alternative data can be used as a scalable and data-driven approach for monitoring firm-level patterns associated with operational deficiencies in consumer-facing financial services.

\begin{table}[htbp]
\centering
\caption{Top 15 Features by Importance}
\label{tab:feature_importance}
\begin{tabular}{lll}
\toprule
Feature & Feature Source & Importance \\
\midrule
BANK OF AMERICA & Structured Categorical Features      & 0.0889 \\
fee/fees           & Text-based       & 0.043 \\
CITIBANK              & Structured Categorical Features      & 0.0324 \\
charge/charges/charged           & Text-based      & 0.0266 \\
bonus           & Text-based      & 0.0248 \\
company primary topic 9   & LDA Topic       & 0.0206 \\
max dollar amount   & Narrative Features Analysis       & 0.0181 \\
Block, Inc.     & Structured Categorical Features     & 0.0178 \\
Chime Financial Inc      & Structured Categorical Features    & 0.0161 \\
company primary topic others          & LDA Topic      & 0.0148 \\
CAPITAL ONE FINANCIAL CORP         & Structured Categorical Features      & 0.0143 \\
dollar amount count  & Narrative Features Analysis      & 0.0126 \\
atm          & Text-based      & 0.0125 \\
TD BANK US HOLDING COMPANY     & Structured Categorical Features      & 0.0118 \\
claim          & Text-based      & 0.0106 \\
\bottomrule
\end{tabular}
\end{table}

\section{Conclusion}
This paper explores whether the outcome involving economic compensation can be predicted based on consumer complaints. We used the complaint data from the Consumer Financial Protection Bureau (CFPB) to construct the prediction task of "whether economic compensation was obtained" as an imbalanced binary classification problem, and developed an analysis framework that combines text features, potential complaint topics, structured category information, and Gradient Boosting algorithms. This method significantly outperforms the benchmark model based on TF-IDF in terms of performance. Its AUC-ROC reaches 0.78, and it still maintains robust performance when evaluated against a wider range of complaint categories.

In addition to predicting performance, research also indicates that consumer complaints contain significant signals related to the outcome of economic compensation. In addition to the text mode, the complaint topic and enterprise-related features also play a positive role in the prediction results of the model.

These research findings highlight the potential of complaint narratives as an alternative source of operational intelligence in the consumer-oriented financial services sector. For regulatory authorities, such models help monitor emerging patterns where consumer rights are infringed upon and determine the priorities of regulatory work. For financial institutions, they help identify recurring service deficiencies and opportunities for operational improvement. From a broader perspective, this research demonstrates how unstructured data generated by consumers can be transformed into actionable insights, thereby simultaneously serving consumer protection and internal learning within institutions.

\end{document}